\documentclass[twocolumn,showpacs,amsmath,amssymb,prb]{revtex4}
\usepackage{graphicx}
\usepackage{dcolumn}
\usepackage{bm}
\begin{document}
\title{Lattice collapse and the magnetic phase diagram of Sr$_{1-x}$Ca$_x$Co$_2$P$_2$}
\author{Shuang Jia$^1$, A. J. Williams$^1$, P. W. Stephens$^2$, R. J. Cava$^1$}
\affiliation{$^1$Department of Chemistry, Princeton University, Princeton, NJ 08544, USA\\
$^2$Department of Physics and Astronomy, Stony Brook University, Stony Brook, NY 11794, USA}

\begin{abstract}

We report that the 122 type Sr$_{1-x}$Ca$_x$Co$_2$P$_2$ solid solution undergoes an anomalous structural transition from the uncollapsed to the collapsed ThCr$_2$Si$_2$ structure at a distinct onset composition near $x=0.5$.
Correlated with the structural changes, the electronic system evolves from a nearly ferromagnetic Fermi liquid to an antiferromagnetic metal, through a complex crossover regime.  The structural collapse, driven by P-P bonding across the (Sr,Ca) layers, is much more pronounced in this system than it is in the analogous Fe-based system, indicating a strong sensitivity of structure to total electron count in the transition metal pnictide 122 family.

\end{abstract}


\maketitle

\section{Introduction}

The layered ThCr$_2$Si$_2$ 122 structure type is commonly observed for AT$_2$X$_2$ compounds based on large (A), transition metal (T), and metalloid (X) atoms.\cite{szytula_handbook_1994}
In this 122 structure, the T$_2$X$_2$ layers, made from edge-sharing TX$_4$ tetrahedra, display a wide range of properties, from magnetic ordering to superconductivity.
Early theoretical investigation of this structure type argued for the critical importance of the shape of the TX$_4$ tetrahedra and X-X bonding across the A layers in determining the electronic states at the Fermi level.\cite{hoffmann_1985}
This has again come to the fore in recent research into the structure-property relationships in the iron pnictide superconductors\cite{Kamihara_Iron, takahashi_superconductivity_2008, Marianne_superconductor}. 
One structural feature of particular interest in the 122 transition metal pnictides is the so-called lattice collapse: some AT$_2$P$_2$ and AT$_2$As$_2$ compounds manifest significantly smaller ratios of stacking to in-plane lattice parameters ($c/a$) than are expected from simple atomic size considerations \cite{Reehuis_structure_1990}. These are called "collapsed tetragonal" (cT) cells, and occur because X-X bonding between T$_2$X$_2$ layers pulls the layers closer and induces a relaxation of the in-plane lattice dimension.
The materials with uncollapsed (ucT) cells are more normal representatives of the structure type, with no X-X bonding present. In some cases the lattice collapse causes a significant difference in Fermi surface topology \cite{analytis-2009, coldea-2009, yildirim_strong_2009}.
CaFe$_2$As$_2$ undergoes a first-order transition from an ucT to a cT phase under pressure.\cite{kreyssig_pressure_2008}

Here we describe the correlations between structure and properties for the Sr$_{1-x}$Ca$_x$Co$_2$P$_2$ solid solution.
The pure Sr and pure Ca end members show a highly anomalous difference in
$c/a$, indicative of a transition from ucT to cT phases\cite{Reehuis_structure_1990, Reehuis_anti_1998}.
Unlike what is expected from simple Vegard's law behavior, here we show that the collapse onsets suddenly in the middle of the solid solution series, even though it is driven by smoothly increasing P-P bonding across the (Sr,Ca) layer.
The magnetic properties of the end member compounds are distinct: SrCo$_2$P$_2$ is a nearly ferromagnetic metal with strongly temperature dependent magnetic susceptibility, whereas CaCo$_2$P$_2$ displays an antiferromagnetic (AFM) transition in which the cobalt moments are ordered ferromagnetically within the basal $ab$ plane but antiferromagnetically along the $c$ axis (A-type AFM).\cite{Reehuis_structure_1990, Reehuis_anti_1998}
Employing diffraction, thermodynamic, magnetic, and transport measurements, we show that the changes in the magnetic properties of the solid solution correlate with the structural anomalies.
The ground states vary from nearly ferromagnetic Fermi liquid (NFFL) to AFM, then to FM-like, and finally back to AFM.
The correlations between the structure and magnetic properties indicate that the electronic structure at the Fermi level for Sr$_{1-x}$Ca$_x$Co$_2$P$_2$ is exceptionally strongly dependent on variations in P-P bonding when compared to other compounds in the same structural family.

\section{Experimental Methods}

Polycrystalline samples were prepared from elemental P, Sr and Ca, and CoP powder\cite{Mcqueen_LaFePO, making}.
All the samples were characterized by laboratory x-ray diffraction (XRD) with $\mathrm{Cu}~K\alpha $ radiation (D8 Focus, Bruker). 
In order to characterize the shapes of the CoP$_4$ tetrahedra and the P-P distances between layers, selected samples, with nominal $x$ equaling 0, 0.2, 0.4, 0.6, 0.7, 0.8, 0.9 and 1.0 were measured by synchrotron powder x-ray diffraction (SXRD) at room temperature at beam line X16C at the National Synchrotron Light Source at Brookhaven National Laboratory.
Structure analysis was performed by using the program GSAS with EXPGUI \cite{toby_expgui_2001, GSAS}.
The refined Ca concentrations $x$ were 1\% - 5\% larger than the nominal $x$.
Therefore, the $x$ values for all samples with the exception of $x=0$ and $1$ were linearly calibrated to the true refined values \cite{Ca}. 
All physical property characterization was performed on a Quantum Design physical property measurement system (PPMS). 

\section{Results}

\begin{figure}
  \begin{center}
  \includegraphics[clip, width=0.45\textwidth]{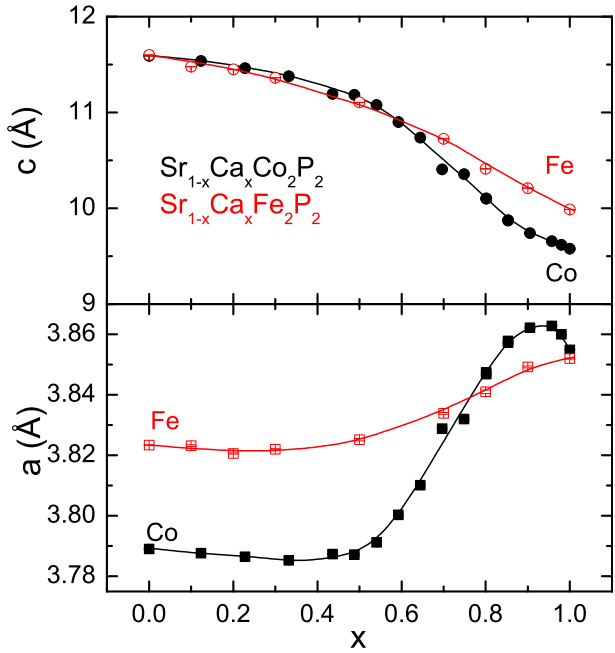}\\
  \caption{(Color online) The lattice parameters for Sr$_{1-x}$Ca$_x$Co$_2$P$_2$ and Sr$_{1-x}$Ca$_x$Fe$_2$P$_2$}
  \label{Fig1}
  \end{center}
\end{figure}

\begin{figure}
  \begin{center}
  \includegraphics[clip, width=0.45\textwidth]{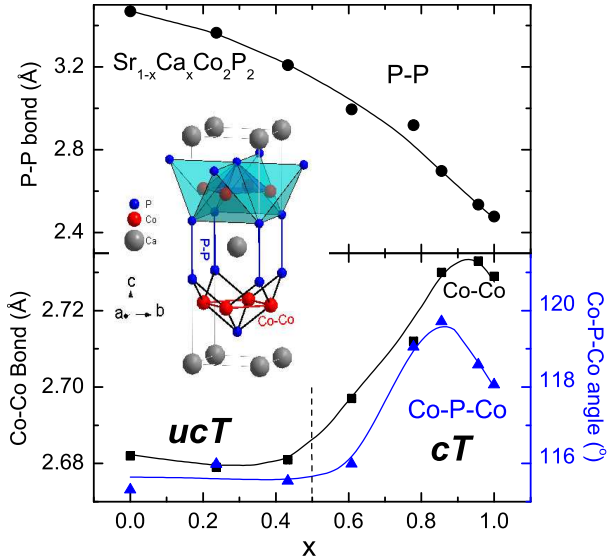}\\
  \caption{(Color online) P-P and Co-Co bond length, as well as the Co-P-Co tetrahedral angle for Sr$_{1-x}$Ca$_x$Co$_2$P$_2$. Inset: the structure of CaCo$_2$P$_2$ unit sell.}
  \label{Fig2}
  \end{center}
\end{figure}

The lattice parameters of the Sr$_{1-x}$Ca$_x$Co$_2$P$_2$ series vary dramatically with composition (Fig. \ref{Fig1}).
The $c$ axis, a measure of the unit cell perpendicular to the T$_2$X$_2$ layers, decreases monotonically, but nonlinearly, with $x$.
The $a$ axis, however, a measure of the T$_2$X$_2$ in-plane dimensions, changes in a highly non-Vegard's-law, $\mathcal{S}$-shape manner with composition, showing minimum and maximum values at $x$ $\sim $ $0.4$ and $0.9$ respectively. There are no two-phase regions in the series; although the onset of the $a$ axis change is sudden, the changes are continuous. The anomalous lattice parameter variation clearly reflects an unusual underlying change in electronic structure that onsets suddenly when proceeding from the ucT (Sr) to the cT (Ca) phases.
The lattice parameters of the Sr$_{1-x}$Ca$_x$Fe$_2$P$_2$ series (made by the same method as the Co samples), by contrast, do not show similarly anomalous variations.
Figure \ref{Fig2} shows the detailed characterization of the crystal structures of Sr$_{1-x}$Ca$_x$Co$_2$P$_2$ series determined by SXRD.
As $x$ increases from 0 to 1, the P-P distance across the Sr$_{1-x}$Ca$_x$ intermediary layer decreases substantially, from 3.3~\AA ~to 2.4~\AA.
The P-P separation changes monotonically with composition. In strong contrast both the Co-Co distance (equaling $a/\sqrt{2}$ and varying from 2.68~\AA ~to 2.74~\AA ) and Co-P-Co tetrahedral angle (varying from $115^{\circ }$ to $121^{\circ }$) vary in an unexpected fashion with composition.
In addition to a distinct onset of a dramatic change at x = 0.5, they display anomalous maxima at $x \sim 0.9$. (Fig. \ref{Fig2}) 

\begin{figure}
  \begin{center}
  \includegraphics[clip, width=0.45\textwidth]{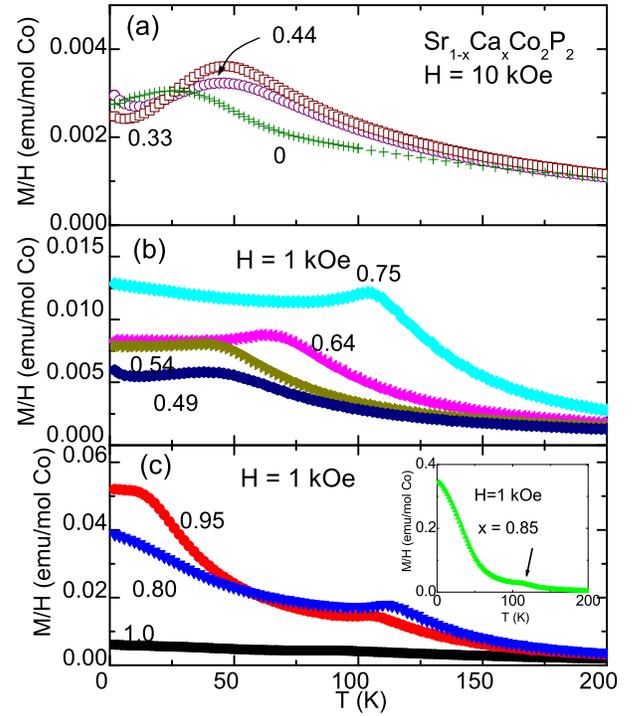}\\
  \caption{(Color online) Temperature dependent $M/H$ for representative members of the Sr$_{1-x}$Ca$_x$Co$_2$P$_2$ solid solution. (a): $x = 0$, $0.33$ and $0.44$; (b): $x = 0.75$-$0.49$ (c): $x = 0.80$, $0.95$ and $1.0$. Inset: $x = 0.85$ (arrow shows the transition point).}
  \label{Fig3}
  \end{center}
\end{figure}

\begin{figure}
  \begin{center}
  \includegraphics[clip, width=0.45\textwidth]{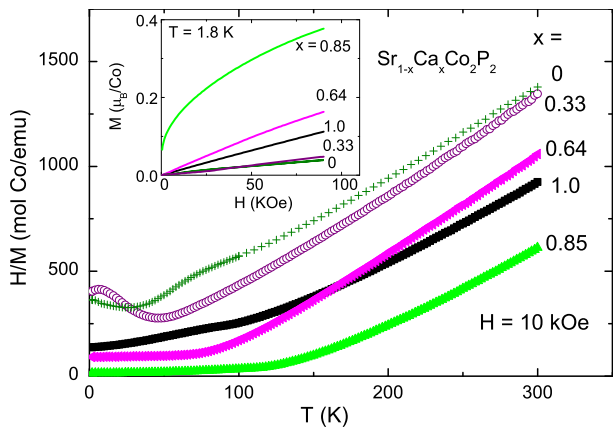}\\
  \caption{(Color online) High-temperature Curie-Weiss behavior for representative members of the Sr$_{1-x}$Ca$_x$Co$_2$P$_2$ solid solution. Inset: $M(H)$ at 1.8~K.}
  \label{Fig4}
  \end{center}
\end{figure}

Figure \ref{Fig3} and \ref{Fig4} presents the magnetic properties for representative members of the series.
The whole series manifests high-temperature Curie-Weiss (CW) behavior ($\chi (T)=C/(T-\theta _{CW})+\chi _0$), with nearly parallel $H/M$ curves (Fig. \ref{Fig4}), indicating similar values of effective moment ($\mu _{eff}$) per Co and differing values of Curie-Weiss temperature ($\theta _{CW}$).
At $x=0$, SrCo$_2$P$_2$ shows an enhanced, temperature-dependent, paramagnetic susceptibility [$\chi (T)$] with a broad maximum (Fig. \ref{Fig3} a) that is typical of nearly FM materials \cite{pd_kia, tibe2_kia, jia_nearly_2007}.
As the Ca content $x$ increases from $0$ to $0.44$, $\chi (T)$ changes very little.

For $x > 0.44$, $\chi (T)$ increases with increasing $x$, and the broad maximum in $\chi (T)$ evolves to a more pronounced feature as $x = 0.54$ (Fig. \ref{Fig3} b).
For $x \geq 0.54$, a sharp maximum in $\chi (T)$ develops, indicating the appearance of an AFM transition. The $M(H)$ curves at 1.8 K in this composition regime are consistent with an AFM ground state (inset of Fig. \ref{Fig3} c).
Both $\chi (T)$ and the AFM ordering temperature ($T_{\mathrm{N}}$) increase as $x$ increases, leading to atypical $\chi (T)$ behavior for $0.80 \leq x \leq 0.95$ (Fig. \ref{Fig3} c and inset of b). 
The $M(H)$ data in this composition regime at 1.8 K (inset of Fig. \ref{Fig4}) show small values of spontaneous magnetization ($ \sim 0.05 \mu _B$/Co for $x = 0.85$), which are much less than the high-field values. This indicates that the magnetic ground state associated with these compositions is somewhat complex -- with a small FM component, rather than being a normal FM or AFM state.
For $x > 0.9$, the $M(H)$ data show no spontaneous magnetization, and $\chi (T)$ decreases dramatically with increasing $x$.
The magnetic ordering temperature also drops with increasing $x$, leading to $T_{\mathrm{N}} = 87 \pm 3$~K for $x = 1$.
This $T_{\mathrm{N}}$ for CaCo$_2$P$_2$ is lower than previously reported (113~K) \cite{Reehuis_anti_1998}, but is consistent with the rest of our series, possibly reflecting a subtle difference of stoichiometry for samples made by different methods.
 
\begin{figure}
  \begin{center}
  \includegraphics[clip, width=0.45\textwidth]{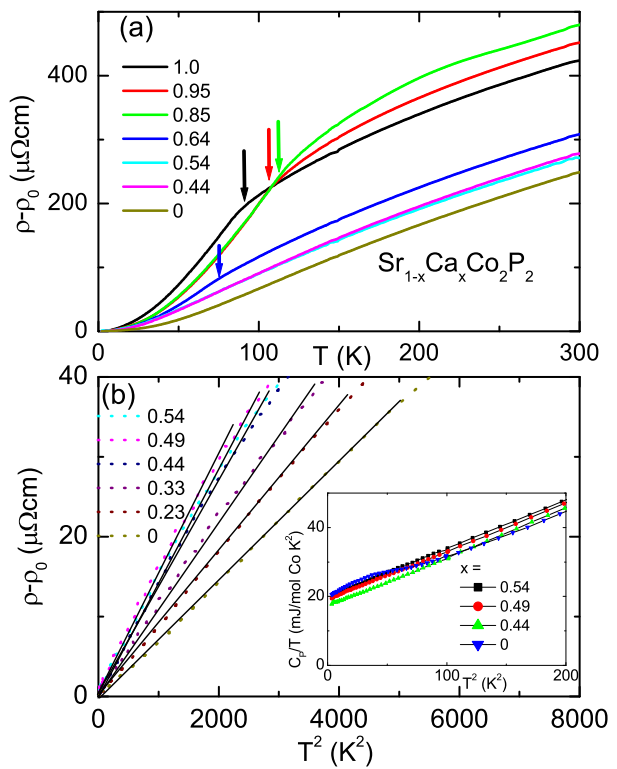}\\
  \caption{(Color online) (a) Temperature dependent resistivity, which has been normalized by the high temperature slope of all resistivity data to that of SrCo$_2$P$_2$ (the arrows show the temperature where the slope changes); (b) low-temperature resistivity versus $T^2$; inset of a: low temperature specific heat for $x \leq 0.54$.}
  \label{Fig5}
  \end{center}
\end{figure}

Figure \ref{Fig5} (a) shows that the temperature-dependent resistivity data manifest a clear slope change due to magnetic ordering for $x \ge 0.64$, but show no anomaly for $x \le 0.54$.
For $x \le 0.54$, the low-temperature resistivity data show FL behavior ($\rho (T) = \rho _0 + AT^2$) below a characteristic temperature (Fig. \ref{Fig5} b).
The $A$ values increase and the characteristic temperatures decrease as $x$ increases.
The low-temperature specific heat data for $x \le 0.54$ (inset of Fig. \ref{Fig5} b) show clear FL behavior ($C_p = \gamma _0 T + \beta T^3$), associated with very similar , intermediate magnitude $\gamma _0$ values ($\sim 20 \mathrm{mJ}/\mathrm{mol Co}\mathrm{K}^2$).

\begin{figure}
  \begin{center}
  \includegraphics[clip, width=0.45\textwidth]{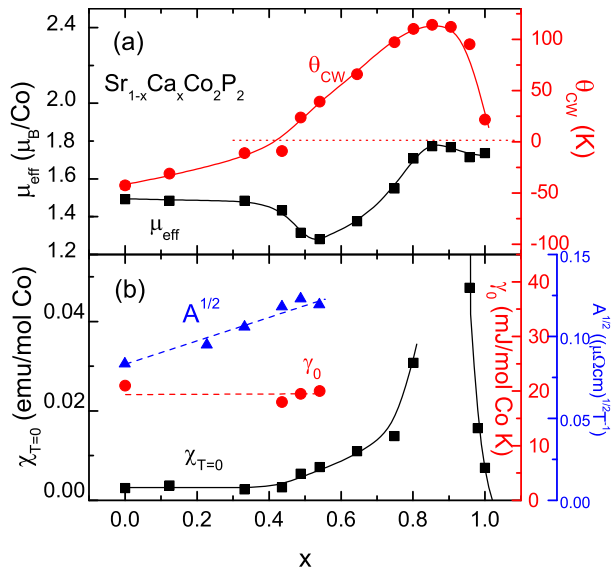}\\
  \caption{(Color online) Summary of physical properties for the Sr$_{1-x}$Ca$_x$Co$_2$P$_2$ series (all lines are guides to the eye). (a) Effective moment ($\mu _{eff}$)and Curie-Weiss temperature ($\theta _{CW}$); (b) zero temperature susceptibility ($\sim M/H$ at 1.8~K), $\gamma _0$ and $A^{1/2}$ for the compositions showing FL behavior}
  \label{Fig6}
  \end{center}
\end{figure}

\begin{figure}
  \begin{center}
  \includegraphics[clip, width=0.45\textwidth]{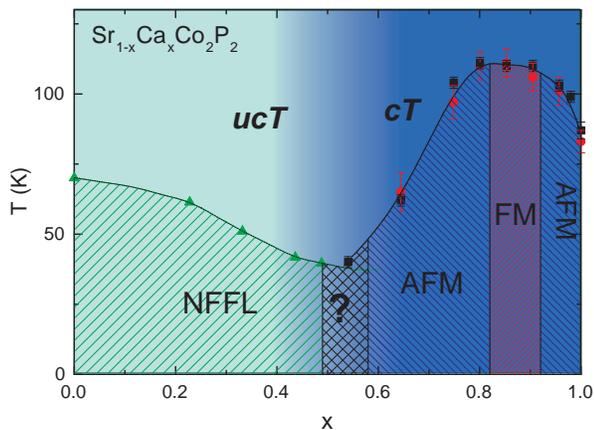}\\
  \caption{(Color online) the electronic and structural phase diagram for Sr$_{1-x}$Ca$_x$Co$_2$P$_2$. ucT and cT: uncollapsed tetragonal and collapsed tetragonal; NFFL: nearly ferromagnetic Fermi liquid; AFM and FM: antiferromagnetic and ferromagnetic order.}
  \label{Fig7}
  \end{center}
\end{figure}

The physical properties of Sr$_{1-x}$Ca$_x$Co$_2$P$_2$ are summarized in Fig. \ref{Fig6} and \ref{Fig7}.
The data show that the magnetic properties for the series are strongly correlated to the variation of the structure (see Fig. \ref{Fig1} and \ref{Fig2}).
The values of $\mu _{eff}$ vary in an $\mathcal{S}$-shape manner, with minimum and maximum values at $x = 0.5$ and $0.85$ respectively. $\theta _{CW}$ varies from negative to positive on going from $x = 0.0$ to $x = 1.0$, showing a crossover from dominantly antiferromagnetic to dominantly ferromagnetic interactions near $x\sim 0.45$, corresponding to the onset of the structural collapse, and showing a maximum value of approximately 100 K at $x\sim 0.8$ (Fig.\ref{Fig6} a).
Figure \ref{Fig6} (b) shows that, as $x$ increases, the values of the zero temperature susceptibility ($\chi _{T=0}$) change relatively little for $0 \geq x \geq 0.45$, and then increase at higher $x$, becoming divergent for $x > 0.8$ where a spontaneous magnetization develops. $\chi _{T=0}$ then decreases again for $x > 0.9$. 
The values of $\gamma _0$ and $A^{1/2}$, which are proportional to the effective mass of the quasi-particles in FL theory, change little for $x < 0.5$, consistent with the behavior of $\chi _{T=0}$.

The electronic and structural phase diagram is summarized in Fig.\ref{Fig7}.
The data show that the system evolves from a NFFL ground state to an AFM ground state through a crossover composition regime near $x = 0.5$. Then, for $0.8 < x \leq 0.9$, the system manifests a FM-like ground state within which the the magnetic ordering temperature is highest near $x = 0.9$. For $x > 0.9$, an AFM ground state reappears.  

\section{Discussion and Conclusion}

The structure and physical property changes in the isoelectronic Sr$_{1-x}$Ca$_x$Co$_2$P$_2$ solid solution are driven by the change in character of the P-P bond across the alkaline earth layer.
As previously described \cite{hoffmann_1985}, the P-P separation in the CaCo$_2$P$_2$ cT phase ($\sim 2.3$ \AA ) is very close to that of a full P-P single bond, yielding an effective electronic configuration of [P-P]$^{-4}$. In the ucT phase SrCo$_2$P$_2$, the P-P distance (3.3 \AA ) is a non-bonding separation, and thus each P can be considered formally as P$^{-3}$.
A transition from a non-bonding to a bonding P-P system as $x$ varies from 0 to 1 can therefore be anticipated, but how that occurs for Sr$_{1-x}$Ca$_x$Co$_2$P$_2$ is surprising.
 
As $x$ first increases, from 0 to 0.4, both $a$ and $c$ decrease slowly, indicating that the slowly increasing P-P hybridization has minimal impact on the electronic system. The replacement of Sr$^{2+}$ by smaller Ca$^{2+}$ in this composition regime can therefore be considered as a simple hydrostatic pressure effect. 
The P-P separation decreases continuously on increasing $x$, and when it reaches a value shorter than 3.1 \AA , near $x = 0.4$, the localization of electrons in the P-P pairs begins to impact the distribution of electrons in the Co$_2$P$_2$ layers, seen dramatically in the changes in the $a$ axis.
The unexpected behavior is that the slowly changing P-P bonding character, as seen in the continuously changing P-P bondlengths, induces a relatively sudden crossover of behavior of the in-plane Co-P electronic system. In this composition regime, there appears to be a sudden onset to the redistribution of charge within those layers in response to the continuously increasing strength of the P-P bond.
The Co-P bondlength changes little across the series (2.23 \AA -2.25 \AA ), indicating that the total charge in the Co$_2$P$_2$ layers is constant.  
For $x$ larger than the other critical value, $0.9$, with the P-P distance shorter than 2.5 \AA , the single P-P bond appears to be fully formed, and both $c$ and $a$ decrease slowly with increasing $x$, again appearing to be a simple chemical pressure effect.
The data in Figure \ref{Fig1} show that the Sr$_{1-x}$Ca$_x$Fe$_2$P$_2$ family does not show a similarly dramatic structural variation.
The absence of this anomaly in the Fe case is interesting, since the P-P distances  are similar to those in the Co system, varying from 3.4 \AA ~to 2.6 \AA , as one goes from Sr to Ca. The difference must therefore be due to the difference in electron count in the T$_2$X$_2$ layer.

Although electronic structure calculations and further experiments are needed to fully understand the phase diagram, some conclusions can be drawn from our observations.
The two-dimensional, characteristic 122 structure of SrCo$_2$P$_2$ indicates that its Stoner enhancement interaction mainly occurs within the Co$_2$P$_2$ layer. This is consistent with the fact that the Co moments in CaCo$_2$P$_2$ ferromagnetically couple within the basal plane. 
For $x < 0.4$, the nearly ferromagnetic FL ground state of Sr$_{1-x}$Ca$_x$Co$_2$P$_2$ is almost invariant (manifested in $\chi _{T=0}$, $\gamma _0$ and $A^{1/2}$), because the P-P distance across the Sr,Ca layer has not reached a critical value at which the P-P bond becomes strong. 
For $x>0.4$, $\mu _{eff}$ decreases, reaches a minimum, and then increases again. 
If the high-temperature CW behavior in these nearly FM compounds is due to spin fluctuations associated with itinerant electrons rather than local moments \cite{moriya_spin_1985}, then the change of $\mu _{eff}$ with $x$ might indicate that itinerant electron spin fluctuations are suppressed and local moments start to form in this composition regime.
This process of local moment formation is correlated with the onset of electron localization in P-P bonds and the resulting redistribution of charge within the CoP$_4$ tetrahedra. This leads to an AFM ground state for $0.6<x<0.8$. Given the positive values of $\theta _{CW}$, this AFM state is presumably A-type. 
The correlation between the magnetic ordering temperature and the Co-P-Co angle and Co-Co distance indicates that both superexchange and direct exchange are important; when the Co-P-Co angle and Co-Co separation reach a maximum, a FM-like ground state appears and the ordering temperature reaches its maximum value.
Although details of the magnetic structure of this FM-like ground state are unknown, AFM to FM transitions strongly correlated to the shape of CoX$_4$ tetrahedra have been seen in other Co compounds in this structure type \cite{huan_co2se2_1989}.
For $x>0.9$, the P-P bond is fully formed and the Co-P-Co angle decreases with $x$, leading to an AFM ground state with slightly lower $T_{\mathrm{N}}$.

In conclusion, our experimental results reveal highly anomalous changes in crystal structure within the Sr$_{1-x}$Ca$_x$Co$_2$P$_2$ series, and correlated magnetic property changes due to the formation of P-P bonds across the Sr,Ca layers that are induced by the substitution of smaller  Ca$^{2+}$  for Sr$^{2+}$.
Due to the continuous nature of the structural changes, the Sr$_{1-x}$Ca$_x$Co$_2$P$_2$ system offers a unique avenue for exploring the evolution of pnictide electronic structures from 2D-like to 3D-like as a consequence of lattice collapse in the 122 structure type.
Further studies such as pressure-dependent magnetic properties and neutron scattering on the compositions in the critical regions, would be of interest.

\begin{acknowledgments}
The authors acknowledge helpful discussions with T.M. McQueen, J.M. Allred, S.L. Bud'ko and J.Q. Yan.
The work at Princeton was supported primarily by the U.S. Department of Energy, Division of Basic Energy Sciences, Grant No. DE-FG02-98ER45706.

\end{acknowledgments}



\begin{thebibliography}{21}
\bibitem[{mak()}]{making}
\bibinfo{note}{Powders were mixed at the ratio of
  $1.1$Sr$_{1-x}$Ca$_x$~:~$2$CoP~:~$0.1$P. Excess Sr, Ca and P were added to
  compensate for high-temperature vaporization. The mixtures were placed in
  alumina crucibles and sealed in evacuated silica tubes and then heated up to
  900 $^\circ $C. The results were then pelletized and heated at 1000 $^\circ
  $C for 12 h. Then the sintered pellets were reground and subsequently pressed
  into pellets, which were sintered at 1100 $^\circ $C for 20 h and then 1150
  $^\circ $C for 20 h.}
\bibitem[{Ca()}]{Ca}
\bibinfo{note}{Actural Ca content ($x_{act}$) was scaled to nominal Ca content
  ($x_{nom}$) by the equation: $x_{act} = 1.04x_{nom}+0.02$.}
\end{thebibliography}
\end{document}